\documentclass[
    amsmath,
    amssymb,
    reprint,
    aps,
    prx,
    superscriptaddress,
    longbibliography
]{revtex4-1}

\usepackage[utf8]{inputenc}
\usepackage[T1]{fontenc}
\usepackage{physics}
\usepackage{nicefrac}
\usepackage{bbm, bm}
\usepackage{makecell}
\usepackage{multirow}
\usepackage{booktabs}
\usepackage{overpic}
\usepackage{tikz}
\usepackage{pgfplots}
\pgfplotsset{compat=1.18}
\usepackage[dvipsnames]{xcolor}

\usepackage[breaklinks=true, colorlinks]{hyperref}
\hypersetup{linkcolor=blue, citecolor=blue, urlcolor=MidnightBlue}
\usepackage{cleveref}

\begin{document}

\title{Absence of quantum advantage for approximate spin glass optimization}

\author{Dries Sels}
\affiliation{Department of Physics, Boston University, 590 Commonwealth Ave., Boston, Massachusetts 02215, USA}
\affiliation{Center for Computational Quantum Physics, Flatiron Institute, 162 5th Avenue, New York, NY 10010, USA}

\author{Flaviano Morone}
\affiliation{Center for Quantum Phenomena, Department of Physics, New York University, New York, NY 10003 USA}

\date{\today}

\begin{abstract}
We perform a semiclassical, large-spin $S$, analysis of the quantum approximate optimization algorithm (QAOA) on the Sherrington-Kirkpatrick (SK) model, using the truncated Wigner approximation. Fixing the QAOA angles to their previously determined optimal $S=1/2$ values, we observe a non-monotonic dependence of the final energy on the spin $S$. At small $S$ the semiclassics is dominated by noise, while the large-$S$ limit is constrained by the exponential growth of the initial fluctuations. For a depth-$p$ QAOA one achieves the optimal balance at $S \sim p$, resulting in a convergence of the final energy to the Parisi value like $\log(p)/p$. We find that the semiclassics slightly outperforms the true spin-1/2 QAOA, and thus suggest they both converge to the Parisi value in the same way. Finally, removing all the initial noise, and re-optimizing the parameters to account for that change, results in superior performance with $1/p$ convergence.  
\end{abstract}

\maketitle

\emph{Introduction --}~Quantum annealing promises to guide complex Hamiltonian systems toward equilibrium faster than thermal annealing, making it a prominent application target for quantum computing~\cite{albash18}. While there exists evidence for polynomial and even exponential quantum speed up for certain problems~\cite{childs03,jordan25}, they typically require a very large amount of qubits and very deep (often exponential) circuits, which are problematic for error correction. The quantum approximate optimization algorithm (QAOA) is designed to try to reduce the requirements on circuit depth~\cite{farhi2014quantumapproximateoptimizationalgorithm}. The QAOA can be viewed as a rudimentary optimal control scheme that attempts to speed up the adiabatic annealing. Although there have been many works on the theoretical and empirical performance of QAOA for optimization problems (see Ref.~\cite{RevModPhys.94.015004} for a review), quantum advantage over the best classical algorithms remains unclear to say the least~\cite{basso2022performance}.

The QAOA has been studied extensively on the Sherrington-Kirkpatrick (SK) model~\cite{Farhi2022quantumapproximate,basso_et_al:LIPIcs.TQC.2022.7,jpmorgan26}, which is a classical spin system with all-to-all couplings between $N$ spins 
and an energy given by
\begin{equation}
    H_c=\sum_{i<j} J_{ij}s^z_i s^z_j,
    \label{eq:HSK}
\end{equation}
where the $J_{ij}$ are i.i.d Gaussian random numbers with zero mean and variance $1/N$. Interestingly, the lowest energy for a typical instance is well understood and reaches the so-called Parisi value $E_P=\left< H_c\right>/N=-0.7631...$ for $N\to\infty$~\cite{PhysRevLett.43.1754, crisanti2002analysis}. 
That this value is well known makes the problem ideal for benchmarking. It should be noted that, even though the problem is hard in the worst case, there are classical message passing algorithms that provably find $1-\epsilon$ approximations to the ground state in $O(N^2)$ time at a cost of scaling exponentially in $1/\epsilon$~\cite{montanari2019optimizationsherringtonkirkpatrickhamiltonian,alaoui2020algorithmic}. There's empirical evidence though that the QAOA would result in a $1-\epsilon$ approximation, with a far better scaling in $1/\epsilon$~\cite{jpmorgan26}. A byproduct of the present paper is that we make a definite prediction for this scaling, which is $\epsilon^{-1}\log \epsilon^{-1}$.

In the context of a recent discussion regarding the beyond-classical nature of a set of spin glass quantum annealing experiments, it was argued in Ref.~\cite{sels2026truncatedwignerdynamicsbiclique} that the universal dynamics of biclique quantum spin glasses -- including sample-to-sample fluctuations of the order parameter-- can be captured by the truncated Wigner approximation (TWA). This raises the natural question how well semiclassical methods, such as TWA, capture the physics of the QAOA. In what follows we report the outcome of a numerical exploration of this question on the SK model. In short, by tuning the effective spin $S$, we can control the amount of quantum noise in the semiclassical simulation and we find it outperforms the exact QAOA at sufficiently large $S$. Further analysis reveals two distinct dynamical regimes, which results in an optimal performance at some intermediate $S^\ast$ which scales with the circuit depth.  
\begin{figure}[t]
    \centering
    \includegraphics[width=0.91\linewidth]{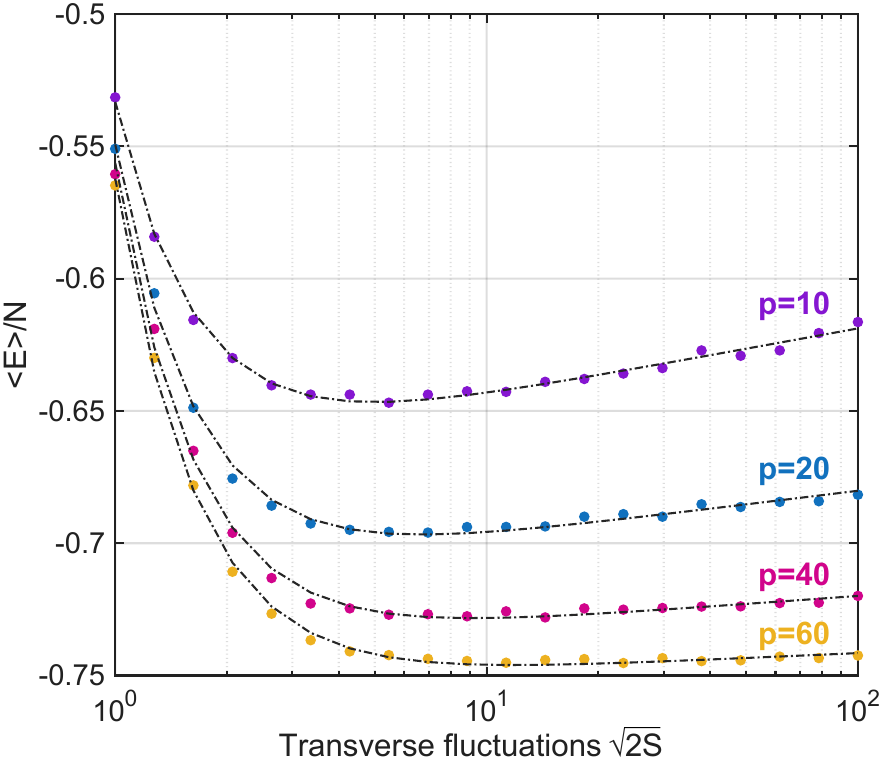}
    \caption{Final energy reached by the semiclassical QAOA as a function of the effective spin, which controls the level of initial noise according to Eq.~\eqref{eq:wigner}. Different depths $p$ are shown, ranging from $p=10$ (purple) to $p=60$ (yellow). The QAOA parameters $\{\gamma\}, \{\beta\}$ are taken directly from Ref.~\cite{jpmorgan26}. Dash-dotted lines show a simple heuristic fit, which capture both the small and large noise limits, i.e. the fitting function is $a/S+b\log(S)+d$ where $a,b,d$ are free parameters, with $a$ tending to a constant at large-$p$ and $b\sim 1/p$.}
    \label{fig:scalingp}
\end{figure}

\emph{Setup --} Consider a depth-$p$ QAOA sequence of alternating unitary transformation generated by the target $H_c$ (given by Eq.~\eqref{eq:HSK}) and rotations around the $x$-axis, yielding the output state: 
\begin{equation}
    \left| \{\gamma\}, \{\beta\} \right>=\prod_{j=1}^p e^{-i \beta_jH_x}e^{-i \gamma_jH_c} \left|+\right>,
    \label{eq:quantumstate}
\end{equation}
with $\left| + \right>$ representing the uniform superposition over computational basis states and $H_x=\sum_i s^x_i$. The angles $\{\gamma\}, \{\beta\}$ are classical parameters that define the QAOA state and they are supposed to be optimized to minimize the expected energy $E_c(\{\gamma\}, \{\beta\})=\langle H_c\rangle$. This optimization 
has been done carefully in Ref.~\cite{Farhi2022quantumapproximate,basso_et_al:LIPIcs.TQC.2022.7,jpmorgan26} and we will use the parameters from Ref.~\cite{jpmorgan26} which were obtained up to $p=80$. 

Instead of considering the exact quantum state in Eq.~\eqref{eq:quantumstate}, we will now restrict ourselves to the semiclassical approximation. In particular, we adopt the truncated Wigner approximation, which is a semiclassical method based on approximating the time-evolution operator in the Weyl-representation~\cite{HILLERY1984121,polkovnikov10,wurtz18}. The method is semiclassical in the sense that it's a saddle-point expansion of a path integral, which for large spin is controlled by $1/S$ serving as an expansion parameter playing the role of $\hbar$, i.e. $1/S$ sets the strength of quantum fluctuations. In practice the TWA amounts to (i) replacing the quantum dynamics 
by classical evolution, and (ii) sampling the initial phase-space points for that classical evolution according the Wigner distribution of the initial quantum state. For spins the equivalent classical dynamics are rotations of the corresponding spin vectors. Consequently, the TWA approximation of the QAOA protocol results in a set of consecutive rotations 
of the spin vectors: each layer applies a $z$-rotation 
of every spin by an angle $\theta^z_i=2\gamma\sum_j J_{ij}s^z_j$ 
set by its instantaneous local field, followed by a 
uniform $x$-rotation by an angle $\theta^x=2\beta$.

The initial phase space points are sampled out of the Wigner function, and since we start from an initial $x-$polarized pure state we can approximate the latter as a Gaussian in the large-$S$ limit~\cite{polkovnikov10}:
\begin{equation}
    W(s_x,s_\perp) \propto e^{-s_\perp^2/S}\delta(s_x-S),
    \label{eq:wigner}
\end{equation}
which indeed captures the correct quantum-mechanical mean and variance of an $x$-polarized spin $S$ state, 
since $\langle s_x\rangle=S$ and $\langle s_\perp^2\rangle=\langle s_y^2+s_z^2\rangle=S$, so that the transverse fluctuations are 
isotropic with variance $S/2$. Note that one could in principle start from the exact Wigner distribution, but this typically does not improve the final result as non-Gaussian corrections are subleading in $1/S$ as compared to the error made by replacing the quantum dynamics with classical dynamics.
In order to not have to rescale the Hamiltonian, it's more convenient to normalize all the $s_x$ to be one, meaning 
we take the initial spins
\begin{equation}
    s_i(0)=(1, s_y,s_z),
    \label{eq:InitialS}
\end{equation}
with $s_{y,z}\sim \mathcal{N}(0,1/2S)$ (the variance is 
reduced from $S/2$ to $1/2S$ because normalizing $s_x\to 1$ 
rescales the transverse components $s_{y,z}$ by $1/S$). 
In practice we sample the initial spins from the 
Wigner distribution of the transverse fluctuations 
in Eq.~\eqref{eq:wigner}, evolve each sample through 
the discrete map 
\begin{equation}
\mathbf{s}_i(n)=R_x[\theta_x(n)]R_z[\theta_i^z(n)]\,\mathbf{s}_i(n-1)\ ,\quad n=1,...,p\ ,
\label{eq:map}
\end{equation}
and average the final energy $H_c[\mathbf{s}(p)]$ 
over the samples and over the couplings. 
It should be noted that, while the approximate large-$S$ Wigner function reproduces the first two moments of the spin distribution, it does so by changing the length of the spins. Since we want to get an accurate estimate of the final energy, it's important to binarize the final spin values. Following Ref.~\cite{morone2026variationaliterativerotationalgorithm}, we propose a simple scheme in which we binarize the final $s_z$ outcomes by aligning them with their final local $z-$field, i.e.
\begin{equation}
    Z_i=-{\rm sign}\left(\sum_j J_{ij}s^z_j(p)\right),
\end{equation}
such that the $Z_i$ are binary random numbers and we can average the final energy $H_c=\sum_{i<j}J_{ij}Z_iZ_j$ over samples, just like in the quantum case. 

\emph{Results--} In figure~\ref{fig:scalingp} we show the 
average energy achieved after a $p$-layer sequence of our semiclassical protocol as a function of the spin $S$. Recall that the latter only controls the amount of noise injected into the initial state, and we're using QAOA angles that are optimized for the exact quantum spin-$1/2$ problem. At small $S$ we observe that the semiclassics fails to reproduce the exact quantum result, producing energies that are significantly higher than what one would get from the respective QAOA. This is not unexpected, since there is no way to reduce the amount of noise during the classical evolution and at small $S$ the initial fluctuations are so large that one is essentially 
starting with a cloud of spins covering half of 
the Bloch sphere. 
In other words, the classical entropy becomes maximal when $S$ tends to zero, resulting in completely random $Z_i$ 
uncorrelated from the couplings and thus $\left<H_c\right> \rightarrow 0$ at small enough $S$. Increasing $S$ reduces the entropy, which means the initial spins start closer to $(1,0,0)$, allowing the $Z_i$ to become correlated with the 
$J_{ij}$ and this lowers the final energy.
However, when $S \rightarrow \infty$ the energy also has to 
vanish, $\left<H_c\right> \rightarrow 0$, because the zero entropy initial state, $s_i=(1,0,0)$, is a fixed point of the classical map~\eqref{eq:map} with zero energy~\cite{morone2026variationaliterativerotationalgorithm}. This implies there has to be a non-monotonic 
dependence of the energy on the spin $S$, with 
an initial decrease of the energy driven by a 
reduction in entropy and a final increase in 
the energy driven by dynamical localization of 
the state around the initial point. 

\begin{figure}[t]
    \centering
    \includegraphics[width=0.91\linewidth]{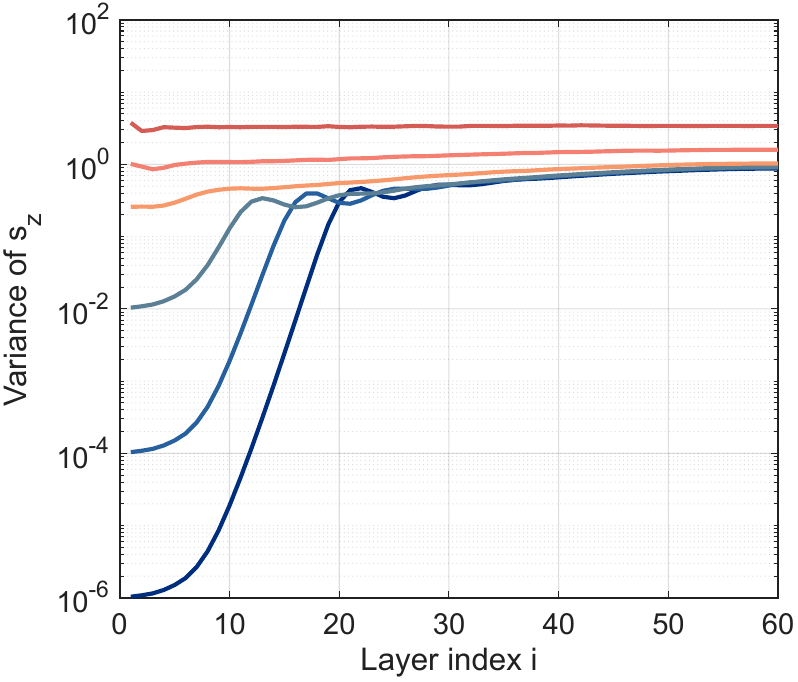}
    \caption{Variance of the semiclassical $z$-magnetization during a $p=60$ QAOA sequence. Different curves show different spin-$S$, which can be identified from the initial variance. For small $S$ (red curves) we observe almost constant variance, while a pronounced regime of exponential growth appears at large $S$ (blue curves).}
    \label{fig:noisetime}
\end{figure}
In the small $S$ regime we find that the 
residual energy above the Parisi value 
$\delta\equiv\langle H_c\rangle/N-E_P$ 
decreases roughly linearly in $1/S$, i.e. 
$\delta\sim a/S$, whereas it increases 
logarithmically with $S$ at large $S$ 
as $\delta\sim b\log(S)$, with $a$ and $b$ 
constant for fixed $p$ (see fit in Fig.~\ref{fig:scalingp}). To understand this, it's instructive to look at the behavior of the $s_z$-fluctuations during the semiclassical evolution, as shown in Fig.~\ref{fig:noisetime}. At small $S$, the fluctuations start out large and remain more or less constant (red curves in Fig.~\ref{fig:noisetime}). At large $S$, when the fluctuations start out small, we observe an exponential increase of the initial fluctuations with time, until they saturate to an (almost) $S$ independent value. The initial exponential growth is simply a consequence of the fact that the $x$-polarized state is an unstable fixed point of the map in Eq.~\eqref{eq:map} for most of the QAOA parameters (see Ref.~\cite{morone2026variationaliterativerotationalgorithm} for a more detailed discussion of this). 
Therefore, the fluctuations grow exponentially at 
a rate set by the Lyapunov exponent of the linearized 
map and this growth stops at the Ehrenfest time $t_E\sim \log(S)$, when the system observes its compactness, i.e. when the initial fluctuations of $O(1/S)$ have grown to $O(1)$.
As long as $t_E \ll p$, one should be able to achieve a low energy, but with increasing $t_E$ there is less time available to actually transport the state to the ground state. 
These two competing mechanisms combine into a single 
model for the residual energy which fits 
the data remarkably well as shown in Fig.~\ref{fig:scalingp}, given by 
\begin{equation}
\delta(S,p)\sim\frac{a}{S} + c\frac{t_E}{p}\sim
\frac{a}{S} +\frac{c\log S}{p}\ .
\label{eq:delta_model}   
\end{equation}
Minimizing over $S$ at fixed depth gives the 
optimum $S^\ast\propto p$, which implies the optimal initial variance $1/2S^*$ scales as $1/p$.

As such, we're guaranteed to have vanishing entropy in the initial distribution when $p\rightarrow \infty$. 
Substituting back yields $\delta(S^\ast,p)\sim c\,\log(p)/p$ where the extra logarithm is simply the 
Ehrenfest time at the optimum, $t_E\sim \log(p)$, 
which grows slowly with depth and degrades the 
naive $1/p$ scaling. 

\begin{figure}[t]
    \centering
    \includegraphics[width=0.95\linewidth]{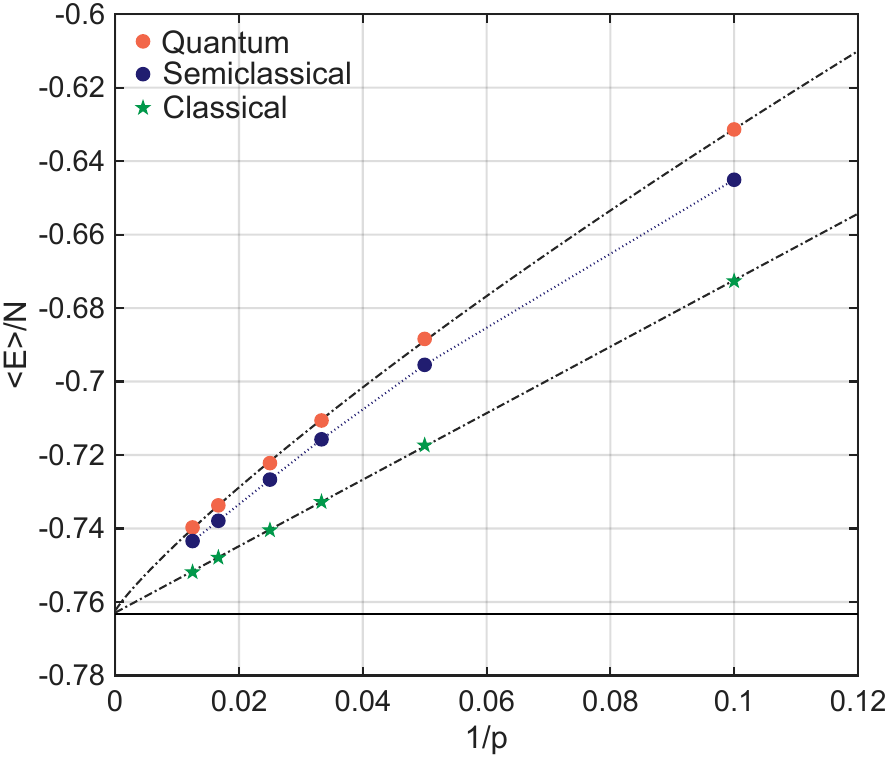}
    \caption{Average energy of the SK model as a function of inverse circuit depth $p$. The quantum results (red dots) are taken from ~Ref.~\cite{jpmorgan26} , we use the same angles for the semiclassical approximation (blue dots) and set the effective spin (noise) to $S=18$. This value is close to the optimum $S^*(p)$ 
    over the displayed $p$-range. Moreover, since the minimum in 
    Fig.~\ref{fig:scalingp} is broad, holding $S=18$ fixed 
    barely increases the energy above its optimum  across $p=10,...,80$.
    The green stars show the fully classical (noiseless) performance after re-optimizing the parameters as discussed in Ref.~\cite{morone2026variationaliterativerotationalgorithm}. The (semi)classical results were run on $N=16384$ and averaged over $10$ samples. The dash-dotted lines show $1/p$ and $\log(p)/p$ fits for the classical and quantum data respectively.}
    \label{fig:Epcompare}
\end{figure}

As shown in Fig.~\ref{fig:Epcompare}, the semiclassical energy is lower than the exact $S=1/2$ QAOA result over the entire range of $p$ (10 to 80) studied here. We have no reason to believe this does not persist to asymptotically large $p$, as long as one keeps increasing $S^*$ accordingly, which implies the QAOA can at best converge to the Parisi value like $\log(p)/p$. In Ref.~\cite{jpmorgan26} a power-law fit of the data suggested a convergence $\approx1/p^{0.88}$, which is consistent with our statement. However, there is no known theoretical reason to justify such a small anomalous exponent, so we argue that the convergence is actually $\log(p)/p$ and it simply appears as power-law because of the small available range in $p$. As shown by the fit in Fig.~\ref{fig:Epcompare}, $\log(p)/p$ convergence is completely consistent with the QAOA data and extrapolates to a value extremely close to the Parisi value.

We consider these observations as compelling evidence for the $\log(p)/p$ convergence of the QAOA to the Parisi value. A number of statements immediately follow. First of all, there is no quantum advantage in using QAOA on SK spin glasses. Secondly, achieving a $1-\epsilon$ approximation to the ground state requires a depth $p=\log (1/\epsilon)/\epsilon$. Each layer requires a matrix-vector multiplication, resulting in an overall complexity of $O\big(N^2 \log (1/\epsilon)/\epsilon\big)$. One might be tempted to argue that the quantum complexity is only $O(N \log (1/\epsilon)/\epsilon)$, but the improvement with $N$ only comes from doing the spin-spin gates in parallel. 
Similarly, one can classically parallelize the 
calculation of the local fields.
Finally, it's clear that the logarithmic slowdown is a consequence of the entropy in the initial distribution. Crucially, there is no need to start from a broad initial distribution as long as one doesn't start in a fixed point of the map~\eqref{eq:map}. In Ref.~\cite{morone2026variationaliterativerotationalgorithm} we present a comprehensive analysis of the scenario in which one can optimize the initial condition and find numerical evidence that the optimal energy converges to the Parisi value like $1/p$, as shown in Fig.~\ref{fig:Epcompare}, thus removing the logarithmic slowdown altogether. 

\medskip

\emph{Acknowledgements}
The Flatiron Institute is a division of the Simons Foundation. D.S. was supported by AFOSR under Award No. FA9550-21-1-0236 and by ONR under Award No. N00014-23-1-2771.

\bibliography{library}

\end{document}